\documentclass[prb,twocolumn,amsmath,amssymb,aps]{revtex4}
\usepackage[dvipdfmx]{graphicx}
\usepackage{dcolumn}
\usepackage{bm}
\usepackage{braket}
\usepackage{hyperref}

\usepackage{ulem}
\usepackage{color}

\newcommand{\red}[1]{\textcolor{red}{#1}}
\newcommand{\blue}[1]{\textcolor{blue}{#1}}

\renewcommand*{\textcolor}[1]{}
\renewcommand*{\sout}[1]{}

\begin{document}
\preprint{APS/123-QED}
\title{\sout{Nuclear magnetic resonance}\blue{NMR}-based gap behavior related to the quantum\sout{-} size effect}
\author{Tomonori Okuno}
\email{okuno.tomonori.77s@st.kyoto-u.ac.jp}
\author{Masahiro Manago}
\author{Shunsaku Kitagawa}
\author{Kenji Ishida}
\email{kishida@scphys.kyoto-u.ac.jp}
 \affiliation{Department of Physics, Graduate School of Science, Kyoto University, Kyoto 606-8502, Japan}
\author{Kohei Kusada}
\author{Hiroshi Kitagawa}
 \affiliation{Department of Chemistry, Graduate School of Science, Kyoto University, 606-8502 Kyoto, Japan
}

\date{\today}

\begin{abstract}
We conducted$^{195}$Pt-nuclear magnetic resonance measurements on various-diameter Pt nanoparticles coated with polyvinylpyrrolidone in order to detect the quantum size effect and the discrete energy levels in the electron density of states, both of which were predicted by Kubo more than 50 years ago. We succeeded in separating the signals arising from the surface and interior regions and found that the nuclear spin-lattice relaxation rates in both regions show the metallic behavior at high temperatures. Surprisingly, the magnetic fluctuations in both regions exhibited anomalous behavior below the same temperature $T^*$, which points to a clear size dependence and is well scaled with $\delta_\mathrm{Kubo}$. These results suggest that a size-tunable metal-insulator transition occurs in the Pt nanoparticles as a result of the Kubo effect.

\end{abstract}

\maketitle



Since the theoretical prediction of the quantum size effect (QSE) on metallic nanoparticles by Kubo~\cite{Kubo1962}, several intensive theoretical studies of this effect have been conducted~\cite{gorkov1965, Sone1977, Halperin1986}.
Kubo predicted that physical properties of metallic nanoparticles would differ from the bulk properties as the system size decreased, and they depend on the parity of an even or odd electron number. For example, the magnetic susceptibilities of nanoparticles with even- and odd-number electrons differ owing to the discrete energy levels~\cite{Kubo1962}.
Kubo determined the mean gap size to be the inverse of the density of states (DOS) $D(\varepsilon_\mathrm{F})$ at the Fermi energy $\varepsilon_\mathrm{F}$ of an entire nanoparticle, considering that the electronic states near $\varepsilon_\mathrm{F}$ in the nanoparticles are redistributed because of the decrease in the particle size.

The experimental results obtained for metallic nanoparticles were summarized in a review article~\cite{Halperin1986}.
Although several studies of nanoparticles have been conducted, the identification of the QSE remains relatively difficult.
This is because the surface effect, which becomes dominant in the small-sized nanoparticles, shows a deviation from the bulk behavior, similar to that expected by the QSE.
The surface regions of nanoparticles often possess non-metallic properties owing to ``a boring surface state,'' e.g., oxidization or degradation by coating materials.
An intriguing QSE obtained from the nuclear magnetic resonance (NMR) measurement of Cu nanoparticle samples has been reported ~\cite{YeeKnightCu, KobayashiCu, GotoCu}.
It was reported that the Knight shift $K$ and nuclear spin-lattice relaxation rate $1/T_1$ exhibited gap behaviors similar to the Kubo effect at low temperatures, although the nanoparticle-size-dependence of the energy gap estimated using $1/T_1$ was much smaller than the theoretically estimated gap, $\delta_\mathrm{Kubo}$.
Unfortunately, surface-interior separation failed because of the small Knight shift of the Cu nanoparticles, owing to the small DOS of the $s$-electron.

In contrast to the nanoparticles of $s$-electron metals, such as Cu, we focused on observing the QSE in nanoparticles of $d$-electron metals, in which the electrons have a large DOS at the Fermi level, to distinguish the QSE from the surface effect.
Although there are many $d$-electron metals, we selected Pt nanoparticles for the following two reasons.
First, $d$-electrons in the Pt nanoparticles are dominant in the DOS at $\varepsilon_\mathrm{F}$ ~\cite{Clogston1964};
thus, surface-interior separation should be easier.
Second, Pt is a good nucleus for NMR measurements~\cite{IUPAC} as the gyromagnetic ratio of $^{195}$Pt is large ($\gamma_\mathrm{n}=9.153$ MHz/T), with a nuclear spin of $I=1/2$, and is thus not subject to the nuclear quadrupole interaction.
Therefore, the shift and linewidth of the $^{195}$Pt-NMR spectrum are determined only through magnetic interaction.
The $d$-electron contribution to magnetic susceptibility can be determined by the Knight shift measurement, as the Knight shift of Pt metal is largely negative owing to the core-polarization effect caused by the $d$-electrons~\cite{Clogston1964}.
Therefore, $^{195}$Pt-NMR measurements of the nanoparticle sample are an excellent means of separately determining the local electronic states at the surface and in the interior regions, as well as distinguishing the QSE from the surface effect.

To date, many NMR studies have been conducted on various-sized Pt nanoparticles~\cite{Rhodes1982-1, Makowka1985, Rhodes1982-2, Bucher1989} to clarify the surface effects but not the QSE.
Rhodes \textit{et al.}~\cite{Rhodes1982-1} observed the $^{195}$Pt-NMR spectra with a broad linewidth at the LN$_2$ temperature; their observations showed a peak in the small Knight shift side. These spectra were assigned to a signalarising from the surfaces of the nanoparticles.
This assignment was directly confirmed using the spin-echo double resonance method~\cite{Makowka1985}.
Bucher \textit{et al}. ~\cite{Bucher1989} pointed out that the NMR spectra are highly dependent on the surface states of the nanoparticles due to the chemisorbtion of oxygen or hydrogen on the Pt nanoparticles.

We measured powdered bulk Pt metal and Pt nanoparticle samples prepared by the reduction of metal ions.
The preparation methods are described in the Supplemental Information~\blue{\cite{SupplementalSamples}}.
The mean diameters of the nanoparticles were 2.5, 4.0, 7.4, and 9.8 nm.
The surfaces of all the samples were covered with PVP to prevent the oxidization and merging of the nanoparticles.
All the samples had a face-centered cubic crystal structure, as determined from an X-ray diffraction pattern measured at room temperature.
The distributions of the particle diameters were evaluated using a transmission electron microscope, as shown in the Supplemental Information~\blue{\cite{SupplementalSamples}}.
For the NMR measurements, 500-mg samples (Pt nanoparticles and PVP) were used.
The NMR measurements were performed using the conventional spin-echo method with a pulsed NMR.
The NMR frequency was fixed to 25.35 MHz, and the NMR spectra were measured using the field-sweep method, as the NMR spectra of the nanoparticles are considerably broader than those of the bulk material.
The fields were calibrated according to the data for the standard Pt bulk embedded in Stycast, and were then converted to a Knight shift.
$T_1$ was measured using the saturation-recovery method at each point on the spectrum i.e. various magnetic fields.
The single-component $T_1$ was evaluated through the exponential fitting in the high-temperature range at which the Korringa relation holds.
At low temperatures, the recovery of the nuclear magnetization shows a multi-exponential behavior, and thus the fitting was performed in two time regions. The fastest and slowest components are shown in Figs.~3 and 4, respectively.
The typical fittings of the recovery of the nuclear-spin magnetization are provided in the Supplemental Information~\blue{\cite{SupplementalRecovery}}.
\red{For the low-temperature measurements, we paid considerable attention to the heating effects by the NMR RF-pulses. As a result, we obtained the NMR spectra and the recoveries of the nuclear magnetization with various-energy RF-pulses.
These are given in the Supplemental Information~\blue{\cite{SupplementalHeating}}.
It was found that the heating effects are negligible for the examined nanoparticles.}

\begin{figure}[tb]
\includegraphics[clip, width= 8.6 cm]{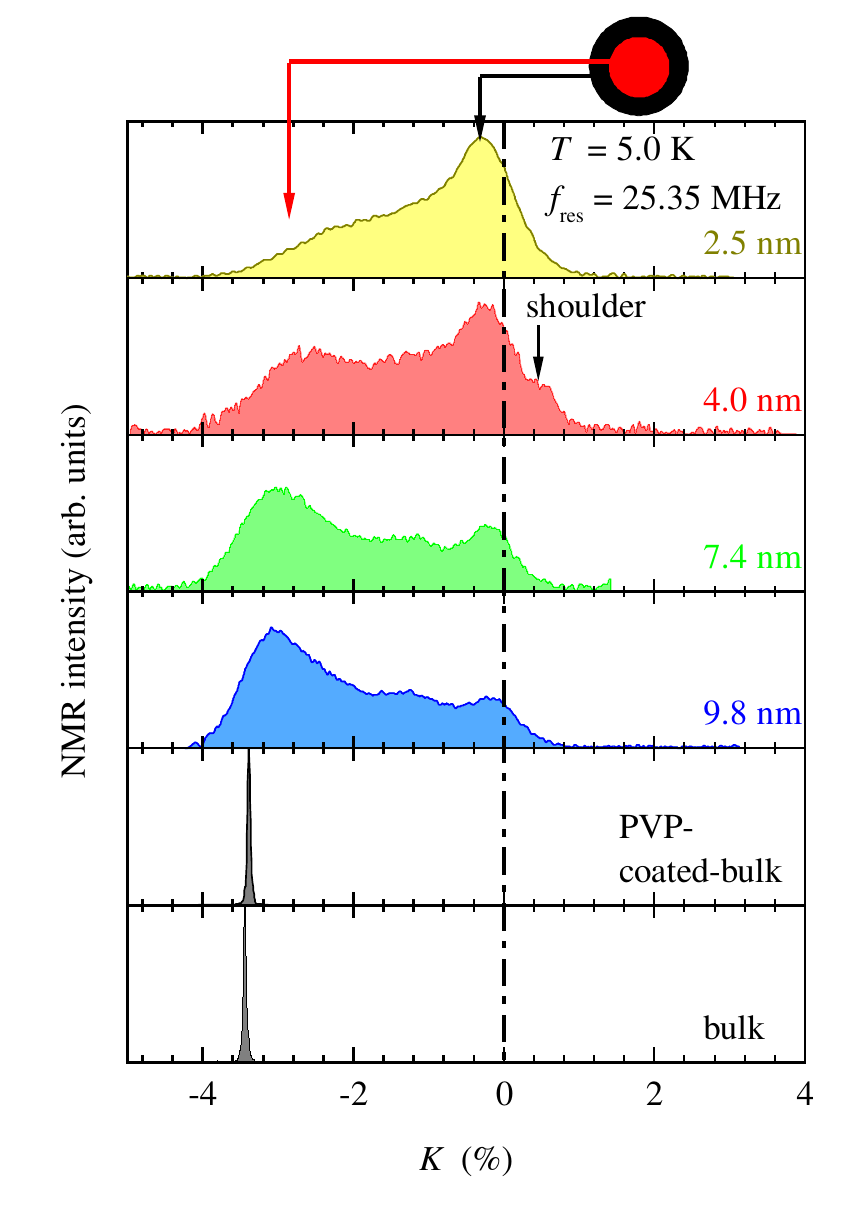}
\caption{\label{fig: NMR spectra} Particle-diameter dependence of $^{195}$Pt-NMR spectra at 5.0 K. The horizontal axis is converted from a magnetic field to a Knight shift. The vertical dotted line represents the position of $K=0$.
In the 4.0-nm sample, a ``shoulder'' structure was observed.}
\end{figure}

The NMR spectra of the samples with different mean diameters (See Sample section) at 5.0 K are shown in Fig.~\ref{fig: NMR spectra}.
A sharp NMR signal was observed in the bulk Pt metal and did not change when a polyvinylpyrrolidone (PVP) coating was used.
With a decrease in the particle diameters, the intensities at $K \sim -3$ \% and $K \sim 0$ \% decreased and increased, respectively.
The particle-diameter dependence of the NMR spectra showed that the signals around $K \sim -3$ \% and $K \sim 0$ \% were assigned as the signals from the interior and surface regions of the nanoparticles, respectively, because the ratio of the number of surface atoms to that of inner atoms increases as the particle diameters decrease.
In the 4.0-nm sample spectra, a shoulder was observed at around $K = +0.46$ \%; this is considered to be the oxidized part of the sample.
In reality, the same signal was observed in the NMR spectrum of the 2.5-nm sample which had been left for a month in the ambient atmosphere, and was assigned as the $^{195}$Pt-NMR signal from H$_2$Pt(OH)$_6$~\cite{Rhodes1982-1, Makowka1985}.
The additional NMR peak~\cite{Rhodes1982-1}, which is often observed on the surfaces of Pt nanoparticles coated with chemical species, was not detected at all, and the present NMR spectra were found to be in good agreement with the NMR spectra of the ``clean'' nanoparticle samples observed in previous studies~\cite{Rhodes1982-1, Bucher1989}; thus, we can conclude that the effect of the PVP coating was negligible.

\begin{figure}[tb]
\includegraphics[clip, width= 8.6 cm]{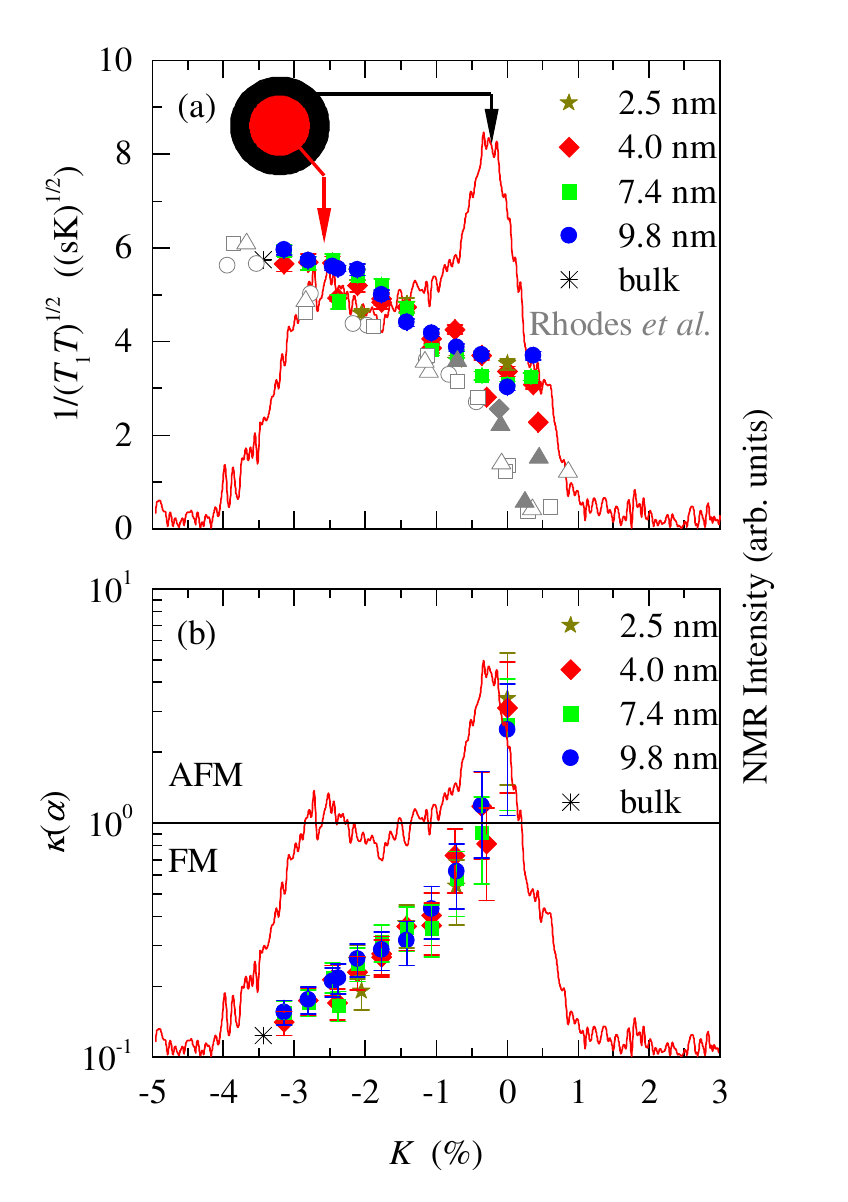}
\caption{\label{fig: T1 position dependence}
(a) Knight-shift dependence of $(1/T_1T)^{1/2}$ at high temperatures, where $1/T_1$ shows the metallic behavior: $T_1T = \mathrm{const.}$ (see Figs.~\ref{fig: 4.0nm T1-T} and \ref{fig: T1-T size}).
The grey points depict the results obtained by Rhodes \textit{et al}.~\cite{Rhodes1982-2}; the symbols are the same as those used in their study.
(b) $\kappa(\alpha)$ is calculated using a modified Korringa relation for each sample.
The error bars indicate the difference in $\kappa(\alpha)$ between $K_\mathrm{orb}$=0.46$\pm$0.2 \%.
The NMR spectrum of the 4.0-nm sample is represented by the red curve.}
\end{figure}

\begin{figure}[tb]
\includegraphics[clip, width= 8.6 cm]{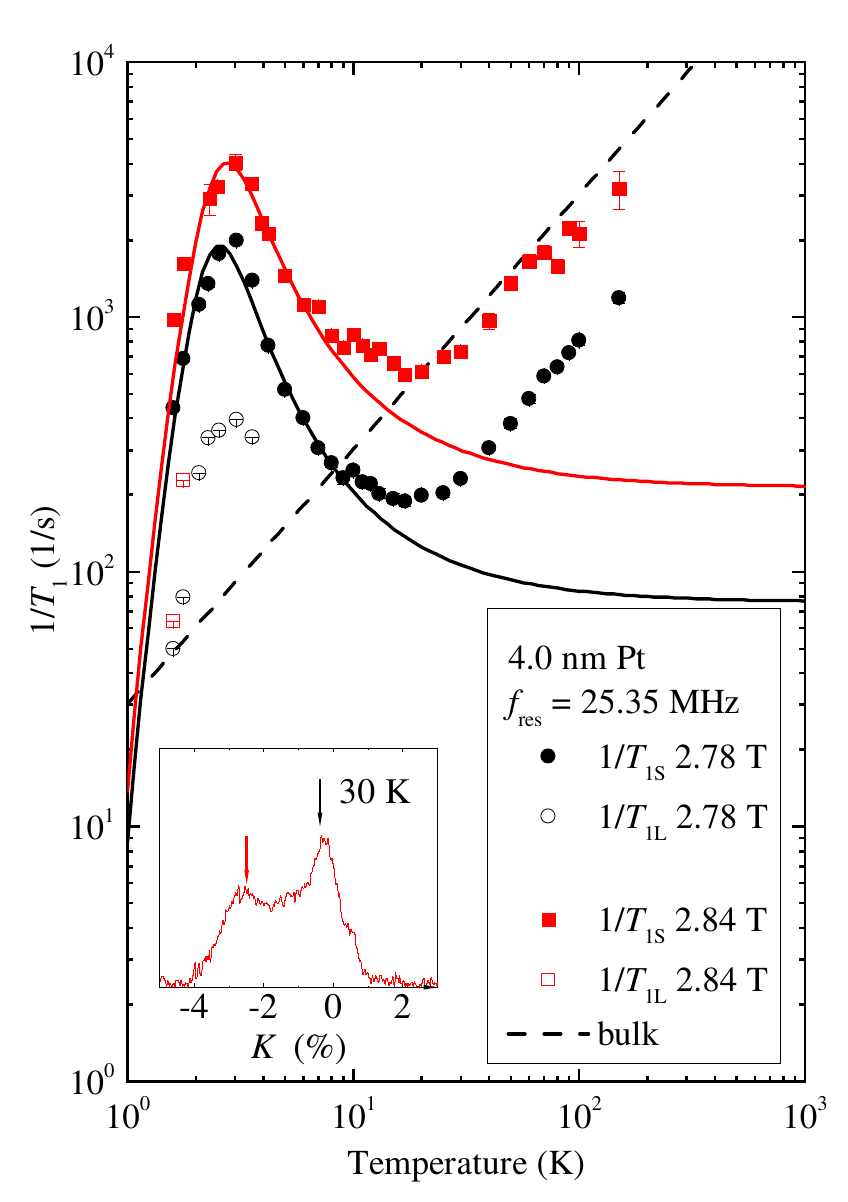}
\caption{\label{fig: 4.0nm T1-T} Temperature-dependence of $1/T_1$ of the 4.0-nm sample. At low temperatures, $1/T_\mathrm{1S (L)}$ is the shortest (longest) component. The solid lines correspond to the values calculated using a BPP model (see text). Inset: The arrows represent the peak positions where $1/T_1$ was measured. The colors of the arrows are the same as those in Fig.~\ref{fig: NMR spectra},~\ref{fig: T1 position dependence} : black and red correspond to signals from the surface and interior of the nanoparticles, respectively.}
\end{figure}

\begin{figure}[tb]
\includegraphics[clip, width= 8.6 cm]{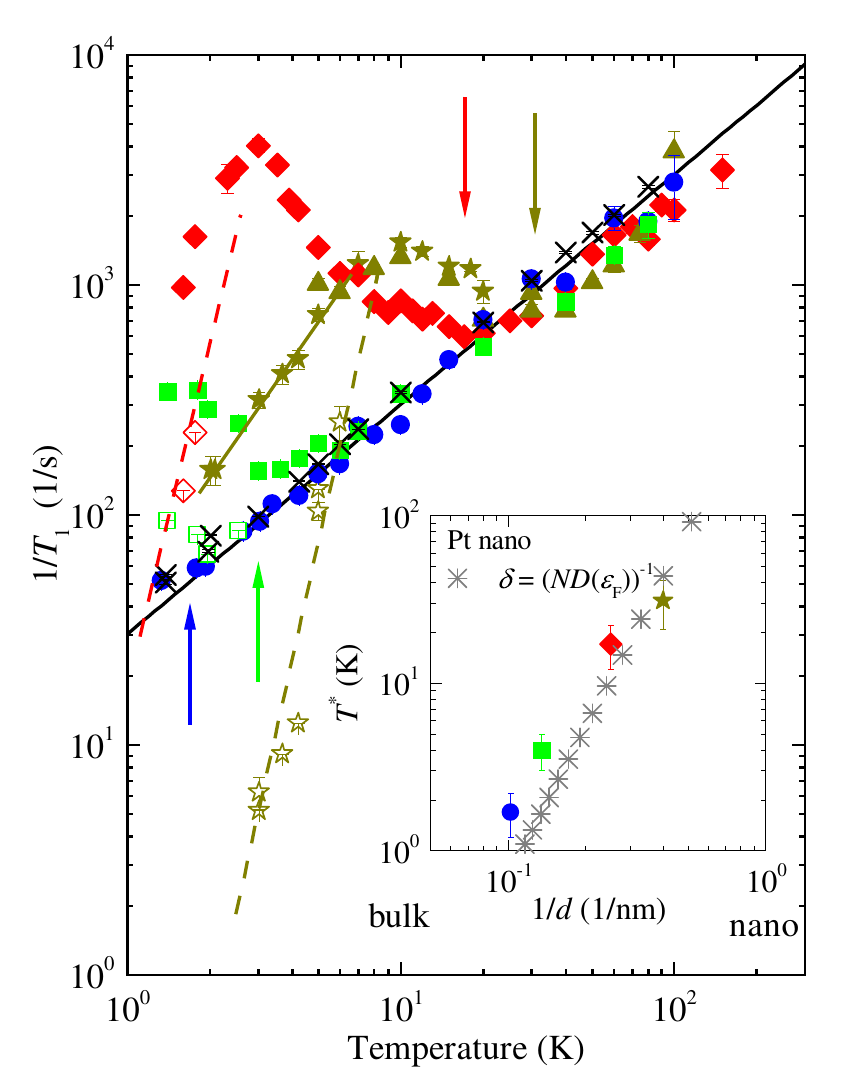}
\caption{\label{fig: T1-T size}Temperature-dependence of $1/T_1$ of each sample, measured at $K \sim -3$ \% and the same magnetic field of 2.84 T.
The yellow $\bigstar$, red $\blacklozenge$, green $\blacksquare$, blue $\bullet$, and black $\times$ symbols show the $1/T_1$ results obtained with the 2.5-, 4.0-, 7.6-, and 9.8-nm nanoparticles and bulk samples, respectively.
The solid and open symbols show the fastest and slowest components, respectively, for the 2.5-, 4.0-, and 7.6-nm samples.
In addition, the arrows show the characteristic temperature $T^*$ of the anomaly.
The solid and dotted lines are guides showing the related data.
Inset: Characteristic temperature $T^*$, where $1/T_1$ deviates from the metallic behavior (represented by the arrows in the main figure).
The horizontal axis shows the inverse of the mean particle diameters $d$ of the nanoparticles.
The calculated value (grey $*$) was obtained from $\delta/k_\mathrm{B} = (k_\mathrm{B} ND(\varepsilon_\mathrm{F}))^{-1}$.}
\end{figure}

To discuss the difference in the electron correlations on the surface and in the interior, the Knight-shift dependence of $(1/T_1T)^{1/2}$ in the ``metallic range'' is shown in Fig.~\ref{fig: T1 position dependence}~(a), which also shows the previous results obtained by Rhodes \textit{et al.}~\cite{Rhodes1982-2}.
The $1/T_1$ results obtained in this study are in good agreement with the results of previous studies~\cite{Rhodes1982-2}.
As discussed below, $1/T_1$ showed the metallic behavior ($T_1T$ = constant) in the high-temperature range, termed the “metallic range” in this paper.
Note that $(1/T_1T)^{1/2}$ and $K$ are related to the local DOS, and $(1/T_1T)^{1/2}$ decreases at the NMR peak with a smaller Knight shift.
More interestingly, the relationship between $(1/T_1T)^{1/2}$ and $K$ is almost independent of the particle diameter.
This clearly indicates that the electronic properties in the metallic range are not characterized based on the particle diameter but are determined through $K$, namely, the local DOS of the 5$d$ electrons.

In general, when the nuclear spin-lattice relaxation is caused by the interaction between the Fermi contact and polarized conduction-electron spins, which also induces the Knight shift, the Knight shift and $1/T_1$ exhibit the following modified Korringa relationship~\cite{Korringa1950, Moriya1962}
\begin{align*}
	\frac{1}{T_1TK^2_\mathrm{spin}}
  &= \frac{4\pi k_\mathrm{B}}{\hbar}
  \left(
    \frac{\gamma_\mathrm{n}}{\gamma_\mathrm{e}}
  \right)^2
  \kappa(\alpha).
\end{align*}
Here, $k_\mathrm{B}$ is the Boltzmann constant, $\hbar$ is the reduced Planck constant, and $\gamma_\mathrm{e}$ and $\gamma_\mathrm{n}$ are the gyromagnetic ratios of the electron and $^{195}$Pt nucleus, respectively.
$\kappa(\alpha)$ is a physical quantity, reflecting the magnetic correlations of the system and originating from the many-body effect. This is a unit for a metal in which electron correlations can be neglected but is much smaller (larger) than unity when ferromagnetic (antiferromagnetic) correlations become significant~\cite{Moriya1962}.
To estimate $\kappa(\alpha)$, we must determine $K_\mathrm{spin}$, as the observed Knight shift is the sum of the spin and orbital contributions, that is, $K = K_\mathrm{spin} + K_\mathrm{orb}$.
Based on the work by Rhodes \textit{et al.}~\cite{Rhodes1982-2}, the shift of the peak of the non-metallic H$_2$Pt(OH)$_6$ was used: $K_\mathrm{orb} = 0.46$ \%.
Further, $\kappa(\alpha)$ was evaluated as shown in Fig.~\ref{fig: T1 position dependence}~(b).
Similar to that of the bulk, $\kappa(\alpha)$ of the interior region is much smaller than unity and increases noticeably from the interior to the surface regions, implying that the electronic correlations differ from the interior to the surface.
The magnetic properties in the interior region are similar to those of the bulk Pt, with the electronic correlation \sout{weakens}\red{become weaker} with a decrease in the local DOS of the $d$-electron near the surface.
\sout{A value of $\kappa(\alpha)$ that is larger than unity at the surface can be explained as follows.
As the local DOS is very small at the surface, the electrostatic screening effect is considered to weaken in this region, thus causing an increase in the electric Coulomb interaction, while the magnetic properties at the surface may be rather close to those of the antiferromagetic state.}
\red{Although the value of $\kappa(\alpha)$ that is larger than unity at the surface might include an ambiguity, the larger value of $\kappa(\alpha)$ of the surface region suggests that the spin correlation might be antiferromagnetic.}
The electronic correlations for one particle seem to be widely distributed; however, the electron correlations were well determined according to the local DOS of the 5$d$ electrons in Pt.

Next, we discuss the low-temperature range, in which anomalous behaviors in $1/T_1$ were observed.
Figure~\ref{fig: 4.0nm T1-T} shows the temperature dependence of $1/T_1$ of the 4.0-nm nanoparticle sample.
The value of $1/T_1$ for both the interior and surface regions exhibited metallic behavior down to 20 K.
Surprisingly, $1/T_1$ for both the regions rapidly increased below 20 K, reached a maximum at approximately 3 K, and then decreased sharply.
The recovery curves of the nuclear magnetization exhibit multi-component behavior below 4 K.
Thus, the fastest and slowest components of $1/T_1$ are shown in Fig.~\ref{fig: 4.0nm T1-T}.
Although the $1/T_1$ values of the interior and surface regions differ, the enhancement behavior of $1/T_1$ is relatively similar between these two regions.
Therefore, this anomalous enhancement can be said to be a result of other than a surface effect and is independent of the local DOS of the 5$d$ electrons for the electronic correlations.
Although this anomalous behavior is reminiscent of the magnetic ordering, no appreciable broadening was observed in the NMR spectra down to 1.5 K; thus, this possibility can be excluded.

Alternatively, we considered that this anomalous behavior may be related to the sample size, and thus investigated the diameter dependence of $1/T_1$ in four nanoparticle samples with different diameters.
Figure~\ref{fig: T1-T size} shows the temperature dependence of $1/T_1$ of the four samples.
In the metallic range, $1/T_1$ was almost independent of the size; however, the broad maximum behavior was observed in all the samples at different temperatures.
The anomalous temperatures, below which $1/T_1$ starts to be enhanced, increase in those samples with a smaller diameter. Thus, we regarded the temperatures at which $1/T_1$ deviates from the metallic behavior (indicated by arrows in Fig.~\ref{fig: T1-T size}) to be the characteristic temperature $T^*$ of this anomaly, and plotted the size-dependence of this characteristic temperature, as shown in the inset in Fig.~\ref{fig: T1-T size}.
The figure also plots the energy-gap temperatures predicted by the Kubo theory,
in which we used the relationship of $T_\mathrm{Kubo} = \delta_\mathrm{Kubo}/k_\mathrm{B} = (k_\mathrm{B}ND(\varepsilon_\mathrm{F}))^{-1}$~\cite{Kubo1962}. $D(\varepsilon_\mathrm{F}) (= 0.853$ states/eV/atom/spin~\cite{Andersen1970}) is the DOS of the bulk Pt, and $N$ is the number of atoms in each sample size of nanoparticles with integer shells.
The characteristic temperatures are in good agreement with the gap temperature estimated using the Kubo theory~\cite{Kubo1962}.

In contrast to the remarkable anomalies in $1/T_1$, we observed no appreciable spectrum change for the Pt nanoparticles at low temperatures.
Several studies~\cite{Taupin1967, YeeKnightCu} have reported that, for Li and Cu nanoparticles, the Knight shift, and thus the spin susceptibility, decreases at low temperatures.
The unchanged spin susceptibility of Pt nanoparticles could be a result of the strong spin-orbit coupling, which suppresses the Kubo effect on magnetic susceptibility~\cite{Sone1977}.
In contrast, as described earlier, a clear anomaly is observed in the temperature dependence of $1/T_1$, which is related to the electronic spin dynamics originating from $\delta_\mathrm{Kubo}$, as spin-orbit coupling generally does not suppress the electron dynamics.

The enhanced behavior of $1/T_1$ can be well interpreted by the Bloembergen—-Purcell--Pound (BPP) model~\cite{BPP1947, BPP2000}.
In this model, the nuclear spin-lattice relaxation is determined by the two-energy level hopping of an electron with the correlation time $\tau$, and $1/T_1$ is expressed as
\begin{align}
  \frac{1}{T_1} = \frac{2\braket{V^2}}{1 + (\omega_\mathrm{n}\tau)^2}
\end{align}
where $\braket{V^2}\sim\braket{(A\delta S/\hbar)^2}$ is a parameter reflecting the strength of the coupling between the nuclear and electronic spins, like hyperfine coupling, and $\omega_\mathrm{n}$ is the NMR resonance frequency.
The solid lines in Fig.~\ref{fig: 4.0nm T1-T} correspond to the fitting of the BPP model with the correlation time following the Arrhenius relation, $\tau = \tau_0 \exp(\Delta/k_\mathrm{B} T)$.
It is not a problem that this model does not explain the behavior in the metallic range, because the BPP model can be effective only for a temperature range that is sufficiently below the gap size, as only two levels are considered.
The parameters used for these are $1/\tau_0=8.0$ (1/ns), $\braket{V^2}=1911\times \omega_\mathrm{n}$ (1/s), and $\Delta/k_\mathrm{B}=9.6$ K for the surface region, and $1/\tau_0=6.1$ (1/ns), $\braket{V^2}=4042\times \omega_\mathrm{n}$ (1/s), and $\Delta/k_\mathrm{B}=10$ K for the interior region.
Although the smaller $\braket{V^2}$ than the Pt metallic hyperfine coupling constant has yet to be explained, it is quite interesting that the gap size $\Delta$ estimated with the BPP model is almost the same as the mean energy gap predicted by the Kubo theory.
This also suggests the presence of the gap at low temperatures, which is related to the Kubo effect.

As $1/T_1$ possibly shows a gap behavior for $1/T_1 \propto e^{-\delta/k_\mathrm{B}T}$, which is far below $T < \delta_\mathrm{Kubo} / k_\mathrm{B}$, the electronic state at low temperatures would be insulating.
Thus, $\delta_\mathrm{Kubo}/k_\mathrm{B}$ may be regarded as being a kind of metal--insulator transition temperature for the nanoparticles and is tunable according to the nanoparticle diameter.
To thoroughly clarify the electronic state at low temperatures, transport measurements using optical probes~\cite{Marquardt1988} have been planned.


In this study, we performed NMR measurements on Pt nanoparticles of various diameters to investigate the electronic states of nanoparticles of $d$-electron systems.
Surface-interior separation was performed, and the difference between the electron correlations of two regions (interior and surface) was clearly shown.
We clarified the anomalous $1/T_1$ behavior at low temperatures originating from the Kubo gap induced by the quantum-size effect; this behavior depends on the nanoparticle diameter and is observed over the entire region of one nanoparticle.
The $1/T_1$ behavior at low temperatures suggests the occurrence of size-tunable metal-insulator transition induced by the Kubo effect.

\subsection*{Acknowledgments}
The authors would like to thank K. Kinjo, G. Nakamine, A. Ikeda, T. Taniguchi, S. Yonezawa, Y. Maeno, and M. Koyama for their valuable discussions and comments.
This work was partially supported by the Kyoto University LTM Center and
Grant-in-Aids for Scientific Research (KAKENHI) (Grants No. JP15H05882, No. JP15H05884, No. JP15K21732, No. JP15H05745, No. JP19K14657, and No. JP19H04696).
We would like to thank Editage (www.editage.jp) for English language editing.



\vspace{1mm}

\begin{figure*}
\includegraphics[clip, bb=75 60 520 780]{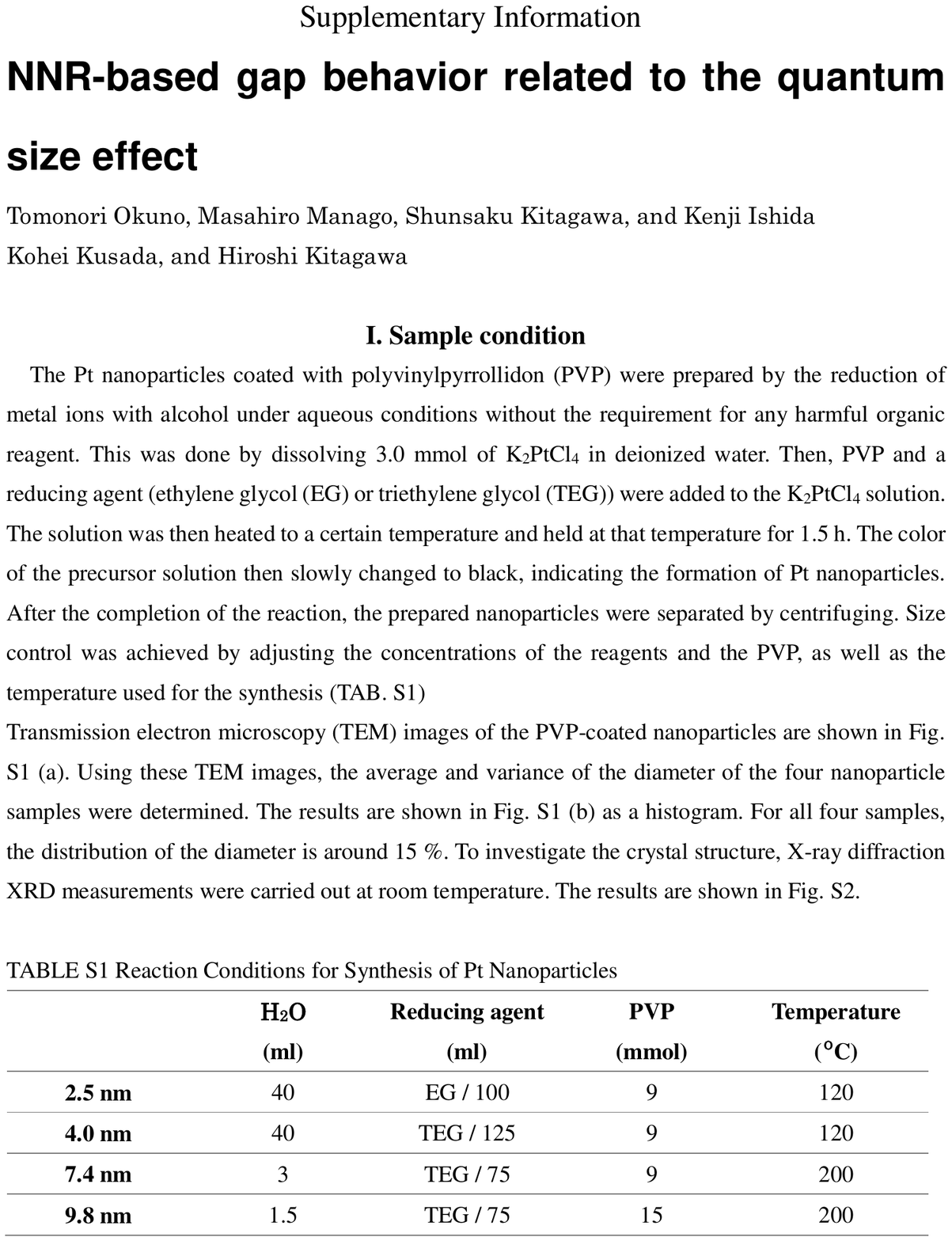}
\end{figure*}
\begin{figure*}
\includegraphics[clip, bb=75 60 520 780, page=2]{./Supplemental.pdf}
\end{figure*}
\begin{figure*}
\includegraphics[clip, bb=75 60 520 780, page=3]{./Supplemental.pdf}
\end{figure*}
\begin{figure*}
\includegraphics[clip,bb=75 60 520 780, page=4]{./Supplemental.pdf}
\end{figure*}
\begin{figure*}
\includegraphics[clip, bb=75 60 520 780, page=5]{./Supplemental.pdf}
\end{figure*}
\begin{figure*}
\includegraphics[clip, bb=75 60 520 780, page=6]{./Supplemental.pdf}
\end{figure*}
\end{document}